\documentclass[aps,prl,twocolumn,superscriptaddress,floatfix,showpacs]{revtex4}

\usepackage{float}
\usepackage{placeins}
\usepackage{amssymb,amsmath}
\usepackage{graphicx}
\usepackage{dcolumn,color,colortbl}
\usepackage[colorlinks=true,linkcolor=blue,citecolor=blue,urlcolor=green]{hyperref}

\begin{document}

\title{Observation of a first-order phase transition deep within the vortex-solid region of YBa$_2$Cu$_3$O$_7$}

\author{M.~Reibelt}
\author{S.~Weyeneth}
\affiliation{Physics Institute, University of Zurich, Winterthurerstrasse 190,
CH-8057 Zurich, Switzerland}
\author{A.~Erb}
\affiliation{Walther Meissner Institute, BAdW, D-85748 Garching, Germany}
\author{A.~Schilling}
\affiliation{Physics Institute, University of Zurich, Winterthurerstrasse 190,
CH-8057 Zurich, Switzerland}
\date{\today}

\begin{abstract}
We have investigated the magnetic phase diagram of a fully oxygenated detwinned YBa$_2$Cu$_3$O$_7$ single crystal by means of magneto-caloric and magnetization measurements, and found thermodynamic evidence for a temperature dependent first-order phase-transition line deep within the vortex-solid region. The associated discontinuities in the entropy are apparently proportional to the magnetic flux density, which may hint at a structural transition of the vortex lattice.
\end{abstract}


\maketitle

The vortex lattice of type-II superconductors in the mixed state can undergo phase transitions. Besides the well known melting transition in cuprate superconductors \cite{Pastoriza1994May,Zeldov1995June,Liang1996January,Welp1996June,Schilling1996August,Roulin1996August,Schilling1997June,Dodgson1998January,Roulin1998February,Bouquet2001May}, other possible kinds of transitions of the vortex lattice have become of growing interest. Nonlocal effects in the vortex-vortex
interaction that are present in certain low-temperature
superconductors such as the borocarbides or V$_3$Si lead
to a rich magnetic phase diagram, with, e.g., structural
phase transitions from cubic to hexagonal vortex lattices
\cite{Kogan1997April,Gurevich2001October}. The structure of the vortex lattice and its phase transitions have been extensively investigated in the past using different imaging techniques \cite{Cribier1964,Essmann1967,Schelten1971,Matsuda1989May,Yethiraj1993,Keimer1994,Maggio-Aprile1995,Shiraishi1998,Johnson1999,Goa2001,Straub2001,Brown2004,Fasano2008,White2009March,Petrovic2009December}. However, thermodynamic data on such structural vortex phase transitions (other than the melting transition) are extremely rare \cite{Park2005February}. The reported first-order transitions in the specific heats of CeCoIn$_5$ \cite{Bianchi2003October} and in the organic superconductor $\kappa$-(BEDT-TTF)$_2$Cu(NCS)$_2$ \cite{Lortz2007November} have been interpreted in terms of a Fulde-Ferrell-Larkin-Ovchinnikov scenario that invokes a vortex segmentation, rather than a mere geometrical rearrangement of the vortex lattice.\\
\indent Very recent small-angle neutron-scattering (SANS) experiments on YBa$_2$Cu$_3$O$_7$ have revealed Fermi surface and/or order-parameter driven, previously unknown structural phase transitions deep within the vortex-solid region, i.e., well below the upper-critical field $H_{c2}(T)$ and the vortex-lattice melting field $H_m (T)$ \cite{White2009March}. At the temperature $T = 2\, $K and with the magnetic field $H$ applied along the $c$-axis, the vortex-lattice structure changes abruptly at $\mu_0 H \approx 2\, $T from a low-field hexagonal to an intermediate-field distorted hexagonal phase, and at $\mu_0 H \approx 6.7\, $T to a high-field rhombic structure. On the basis of the observation of discontinuous changes in the lattice structure, these transitions have been suggested to be of first order and the transition line was found to be temperature independent \cite{White2009March}.\\
\indent In this work we report on the observation of a temperature dependent first-order phase transition line deep within the vortex-solid region. We conducted measurements of the quasi-isothermal magneto-caloric effect in an ultrapure YBa$_2$Cu$_3$O$_7$ single crystal \cite{Erb1996} that is part of the mosaic studied in these SANS experiments. The crystal has a critical temperature $T_c \approx 86\, $K and in a magnetic field of $7\, $T the melting of the vortex lattice was observed at $T_m(7T) \approx 80\, $K \cite{ForganPrivateCommunication}. To reduce vortex pinning on defects, the crystals had been slightly overdoped to achieve full stoichiometry, and they were also mechanically detwinned \cite{White2009March,Hinkov2004August}. The crystal used in our experiments (mass $\approx 6.5\, $mg) was aligned with its $c$-axis parallel to the external magnetic field. The calorimetric measurements were done in a home-built differential-thermal analysis (DTA) setup \cite{Schilling2007March} and run in a constant-temperature mode with varying magnetic field. In experiments of this type, the sample and its suspension with total heat capacity $C$ are connected to a thermal bath with a known heat link $k$ that defines the time constant $\tau = C/k$ according to which the sample can approach the temperature of the thermal bath. On varying the external magnetic field from a starting value $H_0$ to $H$ at a rate $dH/dt$, a deviation $\Delta T(H)$ of the sample temperature from the constant thermal bath temperature $T$ evolves that can be used to calculate the variation in the entropy $S(H)$ according to

\begin{eqnarray} \label{eq.1}
[S(H)-S(H_0)] \frac{T}{C} \approx & - & [\Delta T(H)-\Delta T(H_0)] \\
                                  & - & \int_{H_0}^{H} \frac{\Delta T(H')}{\tau\, dH/dt} dH' \quad , \nonumber
\end{eqnarray}

\noindent which is valid for $\Delta T \ll T$, i.e., for small variations
of $S$ as compared to $C$ \cite{Schilling2007March}. While the first term in equation \eqref{eq.1} originates from the definition of the entropy \emph{S}, the second one represents a deconvolution term accounting for the effect of a finite heat link on the sample temperature in such an experiment, and it vanishes under perfectly adiabatic conditions, i.e., $k = 0$.\\

\begin{figure}
\centering
\includegraphics[width=85mm,totalheight=200mm,keepaspectratio]{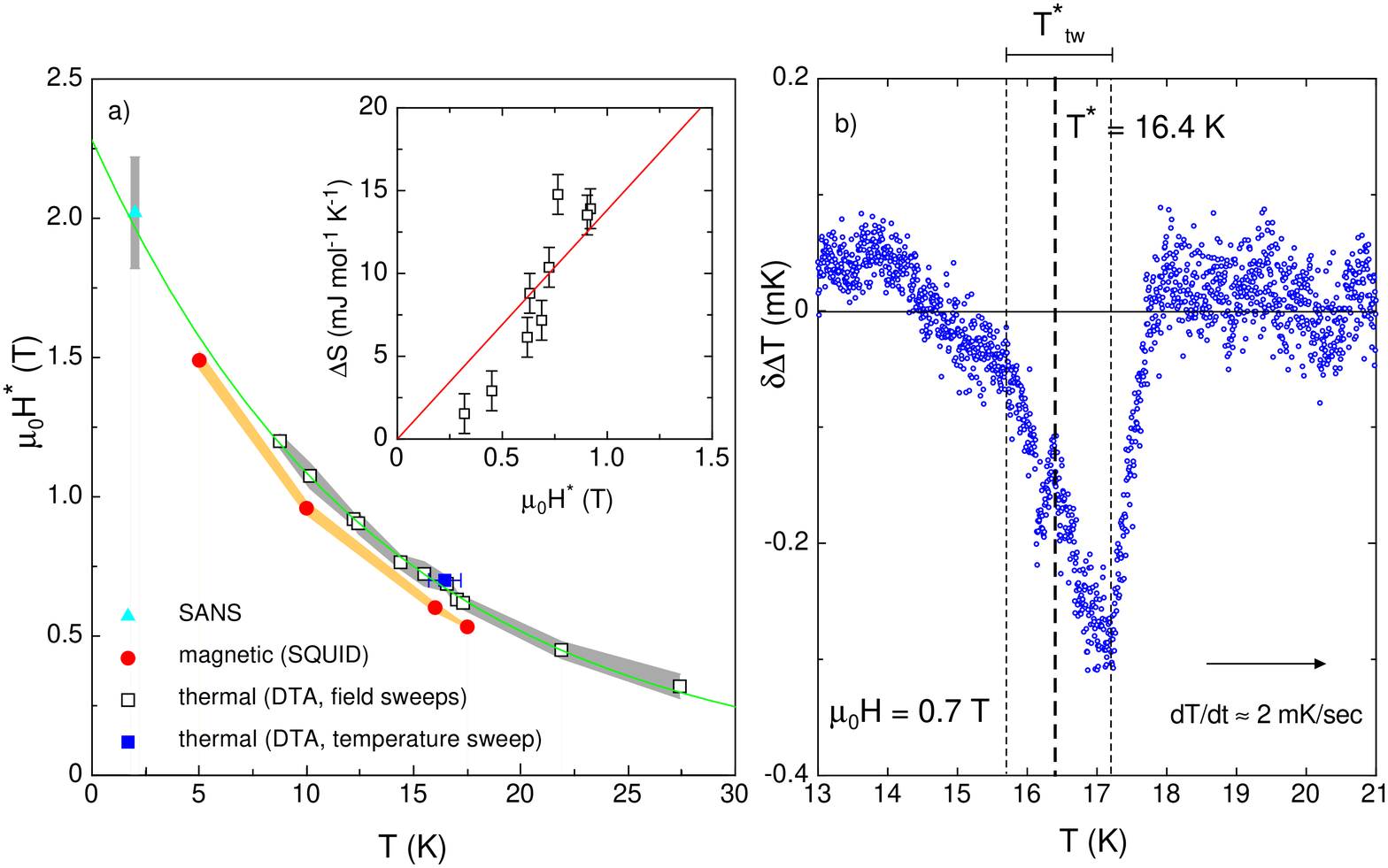}
\caption{\textbf{(a)} Phase diagram of a detwinned stoichiometric YBa$_2$Cu$_3$O$_7$ single crystal as derived from magneto-caloric effect and magnetization data. The SANS data point from \cite{White2009March} has been added to the phase diagram for comparison. The line is an exponential fit to the magneto-caloric data (see the text). The grey and orange shaded areas are derived from the transition widths of the
magneto-caloric and magnetization data, respectively.  Inset: entropy difference between the higher-field and the low-field phase. The line is a linear fit. \textbf{(b)} DTA temperature scan in a constant magnetic field $\mu_0 H = 0.7\, $T after subtraction of a polynomial background. The data are showing the cooling effect due to the latent heat absorbed at the first-order phase transition.
\label{fig_2}}
\end{figure}

\begin{figure}
\centering
\includegraphics[width=85mm,totalheight=200mm,keepaspectratio]{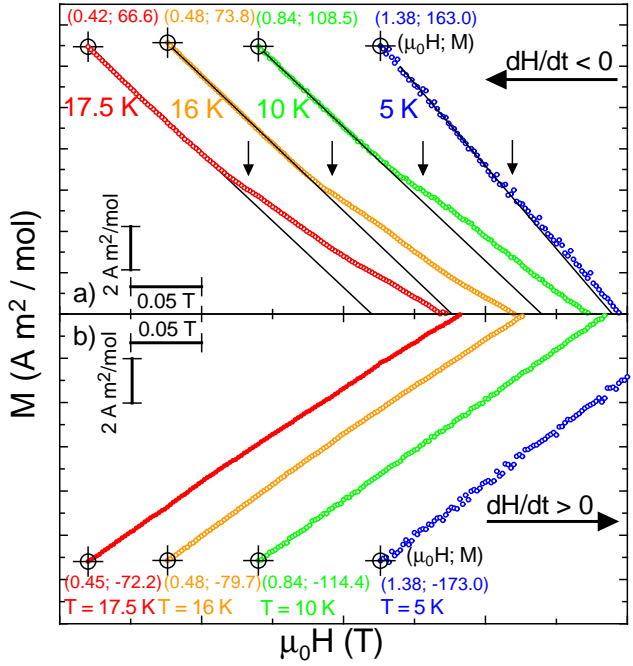}
\caption{Magnetization $M(H)$ of the investigated YBa$_2$Cu$_3$O$_7$ single crystal for \textbf{(a)} decreasing and \textbf{(b)} increasing magnetic field $H \parallel c$ for different temperatures. To make the region of the proposed phase transition at $H^*$ clearer (see the arrows), we have shifted the curves horizontally and vertically on the same relative scales, and indicated the absolute values of $\mu_0 H$ and $M$ for each $M(H)$ curve at a selected data point. The straight lines are fits to the $M(H)$ data for $H<H^*$.}
\label{fig_3}
\end{figure}

\indent White \emph{et al.}~\cite{White2009March} prepared the vortex lattice for the above cited SANS measurements according to two different techniques. The first one was a standard field cooling through $T_c$ in a constant magnetic field (`field cooling', FC). The second one was a field cooling with the main (longitudinal) field oscillating around the target value (`oscillation field cooling', OFC) with an oscillation amplitude of the order of a few mT, and at a frequency $f \approx 30\, $mHz. For both FC and OFC, the field was held stationary for diffraction measurements at $T=2\, K$. The resulting SANS patterns obtained from OFC measurements showed a significant improvement of the vortex-lattice perfection as compared to measurements conducted without using an oscillating field.\\
\begin{figure}[h!]
\centering
\includegraphics[width=85mm,totalheight=400mm,keepaspectratio]{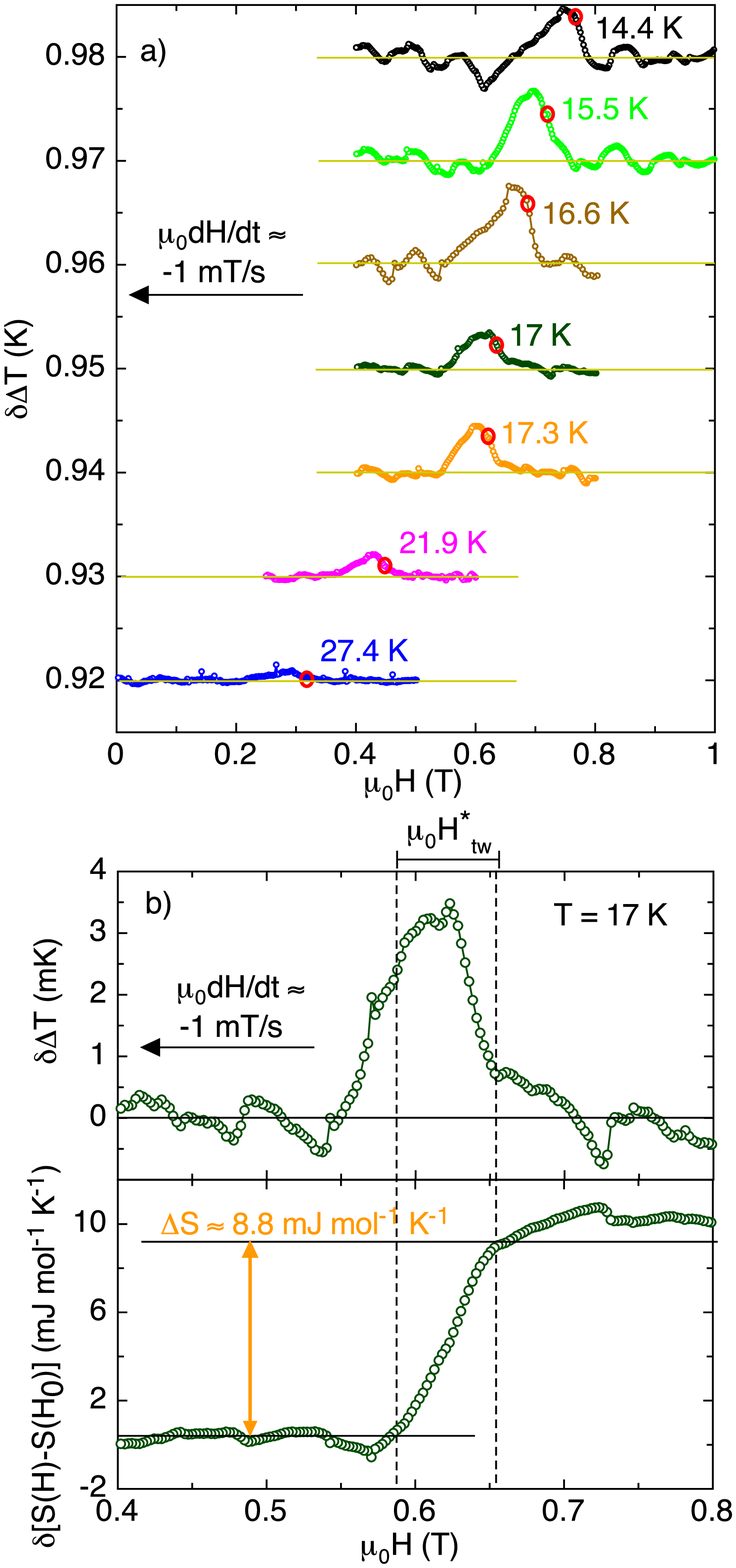}
\caption{\textbf{(a)} Magneto-caloric effect measured in an YBa$_2$Cu$_3$O$_7$ single crystal on ramping the magnetic field $H \parallel c$ down. A polynomial background has been subtracted to obtain $\delta \Delta T$, and the curves are vertically shifted for clarity. The large red circles mark the field which we defined as the transition field $H^*$. \textbf{(b)} Example illustrating the procedure used to extract the changes in the entropy according to \cite{Schilling2007March}.
\label{fig_1}}
\end{figure}
\indent In most of our measurements we cooled the sample to the target temperature in a constant magnetic field $\mu_0H = 7\, $T prior to a measurement. The magneto-caloric effect was then monitored while decreasing $H$ to zero at a constant rate $\mu_0 dH/dt \approx -1\, $ mTs$^{-1}$. Since our magnet power supply is not able to ramp the magnetic field perfectly linear, an ac magnetic field component parallel to the main magnetic field was always superimposed in such experiments with an amplitude $\mu_0 h_{ac} \approx 0.4\, $mT and a frequency $f \approx 11\, $mHz, resulting in conditions very similar to those for the OFC technique used in \cite{White2009March}. According to \cite{Mikitik2001August} the magnetic field of full penetration is $B_p=\mu_0 j_c w$, where $2w$ is the width of the sample and $j_c$ the critical-current density. With $w \approx 0.75\, $mm and $j_c \approx 2.88 \times 10^9\, $Am$^{-2}$ at $T=10\, $K that we estimated from our magnetization data, we obtain $B_p \approx 2.7\, $T, which is much larger than the oscillation amplitudes used in both the SANS study and in our experiments, so no full penetration of the ac magnetic field can be expected.\\
\indent In figure \ref{fig_1}(a) we show the resulting magneto-caloric effect for various temperatures on ramping the magnetic field $H \parallel c$ down (i.e., with $H_0 > H$ and $dH/dt < 0$ in equation \eqref{eq.1}). (The $\delta \Delta T$ data represent the $\Delta T$ data after subtracting a polynomial background.) Abrupt increases in $\Delta T(H)$ at certain magnetic fields $H^*(T)$ indicate the occurrence of a first-order phase transition separating a higher-field phase from a low-field phase with lower entropy \cite{Schilling2007March}. These $\Delta T(H)$ data can be deconvoluted according to equation \eqref{eq.1}, and an example of the resulting background corrected entropy change is shown in figure \ref{fig_1}(b). In figure \ref{fig_2}(a) we have plotted the corresponding magnetic phase diagram, together with the extracted discontinuities in the entropy $\Delta S$ in an inset to that figure. The experiments below $10\, $K were too noisy for extracting reliable $\Delta S$ values due to unwanted thermal effects induced by the superimposed ac magnetic field, although the transition at $H^*$ was still clearly discernible also in these $\Delta T (H)$ data.\\
\indent If the reported features were of thermodynamic origin they ought to be seen also in experiments in the opposite direction, i.e., when crossing the supposed phase boundary $H^*(T)$ from the low-field to the higher-field phase. In related experiments in a constant magnetic field (i.e., without any oscillating field component) but with increasing temperature, we did indeed observe a corresponding cooling effect on crossing the phase-transition line $H^*(T)$ from below (see figure \ref{fig_2}(b)). A corresponding data point is also drawn in the phase diagram in figure \ref{fig_2}(a). Note that the magnitude of $\delta \Delta T$ in the latter type of measurement (figure \ref{fig_2}(b)) is smaller than that measured in the field-sweep experiments shown in figure \ref{fig_1}, which is a direct consequence of the relatively low measuring speed with which the phase-transition line has been crossed. According to \cite{Schilling2007March}, corresponding signals involving the same latent heat $T \Delta S$ scale with the inverse of the time needed to pass a phase transition of finite width (see below), which was $\approx 750\, $s for the temperature scan and $\approx 72\, $s for the field scan at approximately the same temperature $T \approx 16.5\, $K. The observation of both heating and cooling effects, depending on the direction in which $H^*(T)$ is crossed, strongly supports our interpretation that the measured variations in sample temperature are in fact related to abrupt changes in the entropy of the magnetic sub-system (i.e., to the release or the absorption of a latent heat to or from the crystal lattice, respectively). They can therefore not simply be caused by dissipative thermal effects, e.g. induced by a possible vortex motion upon the rearrangement of the vortex lattice, that would lead to a heating effect in all cases.\\
\indent In figure \ref{fig_3}(a) we show magnetization data $M(H)$ measured using a SQUID magnetometer (Quantum Design Inc.) for the same single crystal as we have used in the thermal experiments. On ramping the magnetic field down, we observed step-like features in $M$ that are reminiscent of a first-order discontinuity $\Delta M$ and occur at magnetic fields which essentially coincide with the phase boundary $H^*(T)$ as obtained from the magneto-caloric experiments (see figure \ref{fig_2}(a)). However, these features are embedded in a background magnetization with gradual and perhaps even discontinuous changes in the slope $dM/dH$ around the proposed phase transition at $H^*$. This fact and the presence of a sizeable irreversible contribution to $M$ (see figure \ref{fig_3}) render a reliable extraction of a $\Delta M$ somewhat difficult as the result may depend on the assumptions to describe the underlying background magnetization. In an attempt to make the background-suppression procedure as unbiased as possible, we have calculated the averaged magnetization $M_{av}(H)$ from $M(H)$ data taken for decreasing and increasing magnetic field, respectively, and we plotted the resulting values in the region of interest in figure \ref{fig_4}. For a tentative comparison with the magneto-caloric data, we have also drawn in estimates for a hypothetical equilibrium discontinuity $\Delta M$ that we obtained from the Clausius-Clapeyron equation, $\Delta S(H^*) = - \Delta M \mu_0 dH^*/dT$, where we used $\Delta S$ values from a linear fit to the thermal $\Delta S(H^*)$ data shown in the inset of figure \ref{fig_2}(a). Although the magnetization is hysteretic and corresponding step-like features are virtually absent in the $M(H)$ data taken for increasing magnetic field (see figure \ref{fig_3}(b)), the agreement of the $\Delta M$ values estimated from the magneto-caloric experiments with the order of magnitude of the measured step-like features in $M_{av}$, as shown in figure \ref{fig_4}, is quite fair, and it may hint at the occurrence of a true, thermodynamic phase transition at $H^*(T)$.\\
\indent For completeness we would like to mention that the field inhomogeneity of the solenoid used, providing the main magnetic field over the used scan length ($4\, $cm) used in the DC mode leads to an unavoidable superimposed longitudinal ac magnetic field of the order of $\mu_0 h_{ac} \sim 1$-$3\, $mT at a leading measuring frequency of $\sim 30\, $mHz that must have been present during all of these magnetization measurements.\\
\indent The width of the phase transition at $T \approx 17\, $K is $\mu_0 H^*_{tw} \approx 70\, $mT as derived from the magneto-caloric field scans shown in figure \ref{fig_1}(b). This is comparable to the region where the step-like features and the slope changes in the $M(H)$ data occur (see the horizontal arrow in figure \ref{fig_4}). It also coincides with the width $T^*_{tw} \approx 1.5\, $K obtained from the temperature scan shown in figure \ref{fig_2}(b) that can be converted to $\mu_0 H^*_{tw} \approx T^*_{tw} \mu_0 dH^*/dT \approx 75\, $mT using the slope $\mu_0 dH^*/dT = -0.05\, $TK$^{-1}$ at this temperature. This good agreement of the transition widths lends support to our assumption that the transition observed in the temperature sweep is the same transition as was observed in the field sweeps.\\
\begin{figure}
\centering
\includegraphics[width=85mm,totalheight=200mm,keepaspectratio]{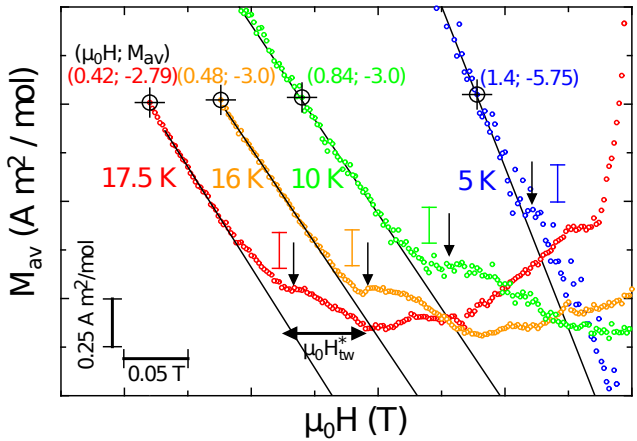}
\caption{Averaged magnetization $M_{av}(H)$ (see the text), shifted in an analogous way to that in figure \ref{fig_3}. Next to the arrows which mark the phase transition at $H^*$  we plotted vertical bars representing estimates for the expected step in the magnetization as derived from our magneto-caloric data and the Clausius-Clapeyron equation (see the text). The horizontal arrow represents the transition width $\mu_0 H^*_{tw}$ as determined from the magneto-caloric measurements (see figure \ref{fig_1}(b)).}
\label{fig_4}
\end{figure}
\indent The phase-transition line, as constructed from our thermodynamic measurements, fairly accurately follows the phenomenological relation $\mu_0 H^*(T) = 2.3 \exp (-T/13.5\, $K$)\, $T and extrapolates at low temperatures to a data point where detailed SANS experiments taken at $T = 2\, $K and in $\mu_0 H \approx 2\, $T have revealed a transformation of the vortex lattice from a low-field hexagonal to an intermediate-field distorted hexagonal phase \cite{White2009March} (see figure \ref{fig_2}(a)). However, additional SANS measurements of the vortex lattice form factor at five different temperatures point to a virtually constant phase-transition field up to the vortex-lattice melting line \cite{WhiteDiss}. This obvious discrepancy to the phase diagram shown in figure \ref{fig_2}(a) may indicate that the phase transition described here is different from that reported in \cite{White2009March} and \cite{WhiteDiss}, and it may therefore also have a different physical origin. The phase transition is probably not a kinetic glass transition from an ordered to a disordered vortex glass since such a transition was recently reported by Petrovi\'{c} \emph{et~al.} \cite{Petrovic2009December} to be not accompanied by a latent heat. A relation to an order-disorder transition in the vortex solid as observed in \cite{Kokkaliaris1999June} is also unlikely since the phase-transition line as derived from our experiments is located at much lower temperatures and magnetic fields in the phase diagram.\\
\indent We want to emphasize that the magneto-caloric measurements probe a bulk property of the ensemble of vortices whose number is given by the flux density $B$, and we therefore judge our magneto-caloric $\Delta S$ data as more reliable than corresponding estimates from the magnetization $\mu_0 M \ll B$, which is highly susceptible to irreversible behavior. It is noteworthy that the measured $\Delta S(H^*)$ scales apparently linearly with the vortex density (see the inset of figure \ref{fig_2}(a)). However, we obtain a $\Delta S \approx 20\, k_B$ per vortex per Cu-O$_2$ double layer (with the Boltzmann constant $k_B$), which is fairly large for a purely structural phase transition of the vortex lattice where one might expect $\Delta S$ to be of the order of $1\, k_B$ per vortex per double layer. Therefore we do not rule out, as already mentioned, the possibility that the observed first-order phase transition may have an entirely different physical origin, and that the apparent agreement of the extrapolated phase-transition line $H^*(T)$ with the SANS data at $T = 2\, $K is coincidental.\\
\indent In summary, we conclude that we have found a thermodynamic evidence for a true first-order phase transition deep within the vortex-solid region of YBa$_2$Cu$_3$O$_7$. The associated discontinuities in the entropy scale virtually linearly with the magnetic field, which may indicate that we observe a structural phase transition of the vortex lattice. Since the transition is of first order, a glass transition can be ruled out as a possible origin. However, the entropy changes are an order of magnitude larger than what has been measured at the known vortex solid-to-liquid transitions, and we can therefore not rule out an entirely different physical origin of the observed phase transition.\\
\indent We thank to E.~M.~Forgan, J.~S.~White, V.~Hinkov, V.~B.~Geshkenbein, and H.~Grundmann for their support and useful discussions. This work was supported by the Schweizerische Nationalfonds zur F\"{o}rderung der Wissenschaftlichen Forschung, Grants Nos.~$20-111653$ and $20-119793$ and the DFG research unit FOR$538$.

\end{document}